\documentclass[twocolumn]{aastex63}

\usepackage{graphicx}
\usepackage{natbib}
\usepackage{multirow}
\usepackage{textcomp}
\usepackage{longtable}
\usepackage{ulem}
\usepackage{soul} 
\usepackage{enumitem}
\usepackage{chngcntr}
\usepackage{hyperref}
\usepackage{changepage}
\usepackage{amsmath}
\newcommand{\bc}{\begin{center}}
\newcommand{\ec}{\end{center}}

\newcommand{\degree}{$^{\circ}$}

\newcommand{\ra}{$\alpha$(J2000)}
\newcommand{\dec}{$\delta$(J2000)}
\newcommand{\h}{$^{\mathrm{h}}$}
\newcommand{\m}{$^{\mathrm{m}}$}
\newcommand{\s}{$^{\mathrm{s}}$}

\newcommand{\kms}{km s$^{-1}$}

\newcommand{\jyb}{Jy beam$^{-1}$}
\newcommand{\jybe}{Jy beam$^{-1}$}

\newcommand{\hii}{H~{\scriptsize II}}

\newcommand{\cyano}{HC$_3$N}
\newcommand{\meth}{CH$_3$OH}

\newcommand{\form}{H$_2$CO}

\newcommand{\sgra}{Sgr A$^*$}

\newcommand{\ie}{i.e.}
\newcommand{\eg}{e.g.}
\newcommand{\til}{$\sim$}

\newcommand{\leftit}{\textit{left}}
\newcommand{\rightit}{\textit{right}}

\begin{document}

\title{\uppercase{6.7 GHz \meth~absorption towards the N3 Galactic Center point-source}} 

\author{Natalie O. Butterfield} 
\affil{Green Bank Observatory, 155 Observatory Rd, PO Box 2, Green Bank, WV 24944, USA}
\email{nbutterf@nrao.edu}

\author{Adam Ginsburg}
\affil{Department of Astronomy, University of Florida, PO Box 112055, Gainesville, FL 32611, USA}

\author{Dominic A. Ludovici}
\affil{Department of Physics and Optical Engineering, Rose-Hulman Institute of Technology, 5500 Wabash Ave, Terre Haute, IN 47803, USA}

\author{Ashley Barnes}
\affil{Argelander Institute f$\ddot{u}$r Astronomy, University of Bonn, Auf dem H$\ddot{u}$gel 71, 53121 Bonn, Germany}

\author{Riley Dunnagan}
\affil{Department of Physics and Optical Engineering, Rose-Hulman Institute of Technology, 5500 Wabash Ave, Terre Haute, IN 47803, USA}

\author{Cornelia C. Lang}
\affil{Department of Physics and Astronomy, University of Iowa, 30 North Dubuque St, Iowa City, IA 52242, USA}

\author{Mark R. Morris}
\affil{Department of Physics and Astronomy, University of California, 430 Portola Plaza, Los Angeles, CA 90095, USA} 

\begin{abstract}

We present evidence of 6.7 GHz methanol (\meth)~and 4.8 GHz formaldehyde (\form) absorption towards the Galactic Center (GC) point-source `N3'. Both absorption features are unresolved and spatially aligned with N3. The 6.7 GHz \meth~contains a single velocity component (centered at \til10 \kms) while the 4.8 GHz \form~shows two velocity components (centered at \til$-$3 and +8 \kms). We find that the velocity of these absorption components are similar to that of emission lines from other molecules (\eg, SiO and \cyano) detected toward this compact-source ($-$13 to +25 \kms; `N3 cloud'). The detection of these absorption features is a firm indication that some of the molecular gas in the N3 molecular cloud is on the near-side of the continuum source.  Analysis of the \meth~absorption kinematics shows a relatively large velocity dispersion (3.8 \kms) for the size scale of this feature \citep[$<$0\farcs1, $<$0.01 pc at the GC;][]{dom16}, when compared with other similar size GC clouds in the Larson linewidth-size relationship. Further, this linewidth is closer to velocity dispersion measurements for size scales of 1.3 pc, which is roughly the width of the N3 cloud (25\arcsec; 1.0 pc). We argue that this relatively broad linewidth, over a small cross-sectional area, is due to turbulence through the depth of the cloud, where the cloud has a presumed line-of-sight thickness of \til1 pc. 

\end{abstract}


\section{Introduction}

The 6.7 GHz \meth~(5$_{1}^{+}$--6$_{0}^{+}$) line is typically observed as a class II maser. Class II masers are radiatively pumped and are therefore frequently associated with star-forming regions \citep[\eg,][]{Menten91b}. While detections of the 6.7 GHz \meth~line in \textit{emission} are common \citep[\eg,][]{Menten91b, Caswell10, rickert18, xing19}, detections in \textit{absorption} are not. As of this paper, only a few studies have reported the 6.7 GHz \meth~line in absorption \citep[\eg,][]{Menten91b, Imp08, Pandian08}. Only a few of the sources in the \cite{Menten91b} survey contained absorption features towards extended sources, all of which are located in the central region of our Galaxy, including: Sgr A-A, Sgr A-F, and the Sgr B2 complex.

\begin{figure*}[tb!]
\centering
\includegraphics[scale=0.645]{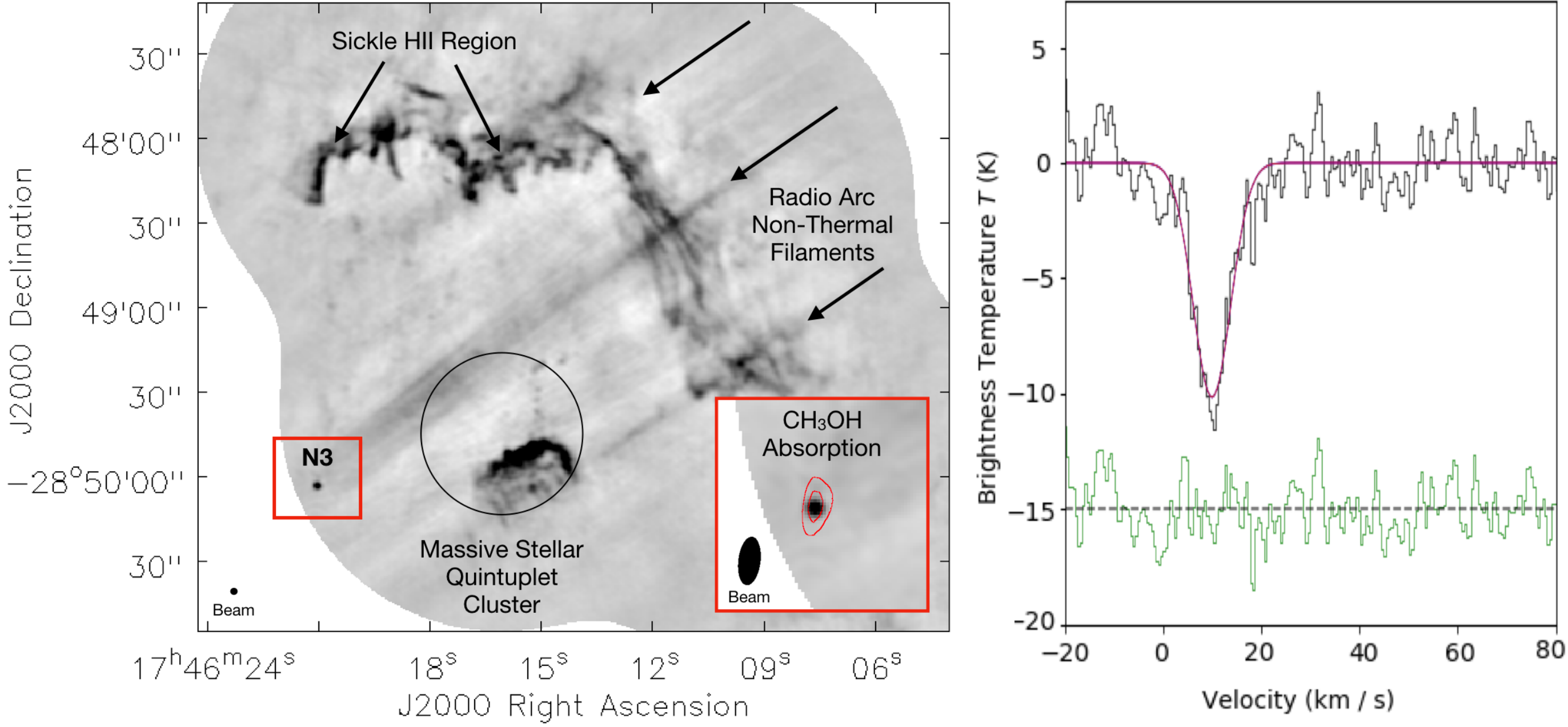} 
 \caption{\leftit: 24.5 GHz VLA radio continuum emission showing the region around the point-source N3 \citep[Figure 2 from][]{my17}.  Several prominent radio continuum features are labeled in the figure. Among them are the Radio Arc Non-Thermal Filaments which are discussed in Section \ref{n3i}.  \textit{inset}: Zoomed in detail of the 6.7 GHz \meth~absorption in the 9.97 \kms~channel shown in red contours at -5 and -10 $\times$ 1.5 m\jyb~(rms noise; Table \ref{params}).  \rightit: 6.7 GHz \meth~absorption (black spectrum) extracted using a region size comparable to the point spread function of the data (see $inset$). The red line shows the best fit parameters presented in Section \ref{abs}, with the green residuals of the fit offset at -15 K. }
\label{finder}
\end{figure*}

The center of our Galaxy (Galactic Center; GC) is typically defined as the inner 500 pc of the Milky Way galaxy.
The environment of molecular gas in the GC is more extreme than in the Galactic disk, with higher gas temperatures ranging from 50$-$300 K \citep[in comparison to 10$-$20 K in the disk; \eg,][]{mills13, Ginsburg16}. Molecular clouds in the GC are also observed to have broader linewidths (2$-$30 \kms~on 5$-$40 pc size-scales) than clouds in the Galactic disk \citep[{0.5$-$5 \kms}; \eg,][]{oka01a, heyer09, shetty12}. The relationship of cloud size to velocity dispersion (also known as the `Larson linewidth-size ($\sigma$$-$$R$) relationship') holds across a wide range of physical size scales from molecular cloud complexes to compact cloud cores \citep[\eg,][]{larson81}. Most of the recent measurements of GC cloud kinematics, across multiple size scales, use dendrogram analysis to fit the velocity dispersion and angular size of the gas components \citep[\eg,][]{shetty12, kauffmann17a}. We employ an alternative method of investigating GC gas kinematics at sub-parsec ($<$0.01 pc) size scales by using a 6.7 GHz \meth~absorption measurement towards the compact source N3.

\subsection{N3 point-source}
\label{n3i}

The `N3' point-source is located 14\arcmin~(33 pc) in projection from \sgra~in a region known as the `Radio Arc' (see Figure \ref{finder}, left).  The N3 point source is spatially projected onto one of the Radio Arc Non-Thermal Filaments \citep[\eg,][]{YZ87}, but appears to be a distinct object (Figure \ref{finder}, left). The nature of the N3 compact source has been a mystery for several decades \citep[\eg,][]{YZ87, dom16, stag19}. A recent study of this compact source determined N3 has a broken power-law spectrum, consistent with synchrotron self-absorption, and is shown to be time-variable over decade-long timescales \citep{dom16}. N3 is unresolved at 44 GHz (\ie, with a 0\farcs26$\times$0\farcs11 beam), the highest spatial resolution in \cite{dom16}, indicating a maximum size of \til1000 AU at a distance of 8.18 kpc \citep{abu19}. Based on the arguments presented in \cite{dom16}, possibilities that N3 is a young supernova, foreground active star, ultracompact \hii~region, or a micro-quasar are all ruled out.\footnote{\cite{dom16} also disfavor N3 being an active galactic nucleus (AGN), but as we discuss in Section \ref{arrange}, that possibility should remain viable. } 

The N3 source has associated molecular emission (the `N3 molecular cloud') that appears to wrap around the compact source \citep{dom16, my17}. The molecular emission is extended, with a diameter of only 25\arcsec~(1.0 pc). The high gas temperatures observed in this cloud \citep[80$-$100 K,][]{dom16} suggest this cloud is located in the GC. Thermal 48 GHz \meth~(1$_0$$-$0$_0$) emission is detected in the N3 molecular cloud along with sixteen 36 GHz \meth~(4$_{-1}$$-$3$_0$) masers and twelve 44 GHz \meth~(7$_0$$-$6$_1$) masers distributed throughout the cloud \citep{dom16}. Both of these \meth~maser transitions are class I masers,\footnote{Class I masers are $collisionally$ excited and therefore trace shocks.} however, only the 36 GHz maser is spatially co-located with the radio continuum of N3.


\section{Observations and Data Reduction} 
\label{obs}

This paper focuses on the 6.7 GHz \meth~transition towards the point source N3. These observations are part of a larger `Karl G. Jansky Very Large Array', hereafter `VLA', survey of the GC (Project ID: 17A-321).\footnote{The VLA radio telescope is operated by the National Radio Astronomy Observatory (NRAO). The National Radio Astronomy Observatory is a facility of the National Science Foundation operated under cooperative agreement by Associated Universities, Inc.} These observations were taken in C band with the C array, resulting in a spatial resolution of \til5\arcsec. We used the Common Astronomy Software Application (CASA)\footnote{\href{url}{http://casa.nrao.edu/}} pipeline, provided by NRAO, to calibrate the data. The spectral line was continuum-subtracted in the UV plane using the CASA task UVCONTSUB before any cleaning was done.  The imaging parameters of the \meth~(5$_{1}^{+}$--6$_{0}^{+}$) data cube are presented in Table \ref{params}. 

\begin{table}[bt!]
\caption{\textbf{Image and Fit Parameters} }
\centering
\begin{tabular}{lc}
\hline\hline
\textbf{Parameter} & \textbf{Value} \\
\hline
\multicolumn{2}{c}{\textbf{Image Parameters}} \\
\hline
\meth~Rest Frequency ~ ~ ~ ~ ~ ~ ~  ~ & 6.66852 GHz \\
Spatial Resolution &  11\arcsec $\times$ 5\arcsec~(0.4$\times$0.2 pc) \\
Spectral Resolution & 0.47 \kms  \\
Channel Noise & 1.5 m\jybe \\
\hline
\multicolumn{2}{c}{\textbf{Fit Parameters}} \\
\hline
Central Velocity & 9.94 $\pm$ 0.17 \kms \\
Velocity Dispersion & 3.84 $\pm$ 0.17 \kms  \\
Depth & -10.15 $\pm$ 0.38 K \\
\hline
\end{tabular}
\label{params}
\end{table}




We also present the detection of 4.8 GHz \form~1(1,0)--1(1,1) absorption towards N3. 4.8 GHz \form~absorption is known to be quite common in the GC \citep[\eg,][]{whiteoak79}. The \form~data presented in this paper were taken using the VLA in C band with the A array (0\farcs4$\times$0\farcs8~resolution) with a sensitivity of 1.0 m\jybe~and velocity resolution of 3.88 \kms~(Project ID: 18A-303). Due to the narrow velocity range, and courser spectral resolution, of the \form~observations, fitting the spectrum is challenging. Therefore, we will be focusing on the 6.7 GHz \meth~transition for the majority of the paper.


\begin{figure}[tb!]
\centering
\includegraphics[scale=0.59]{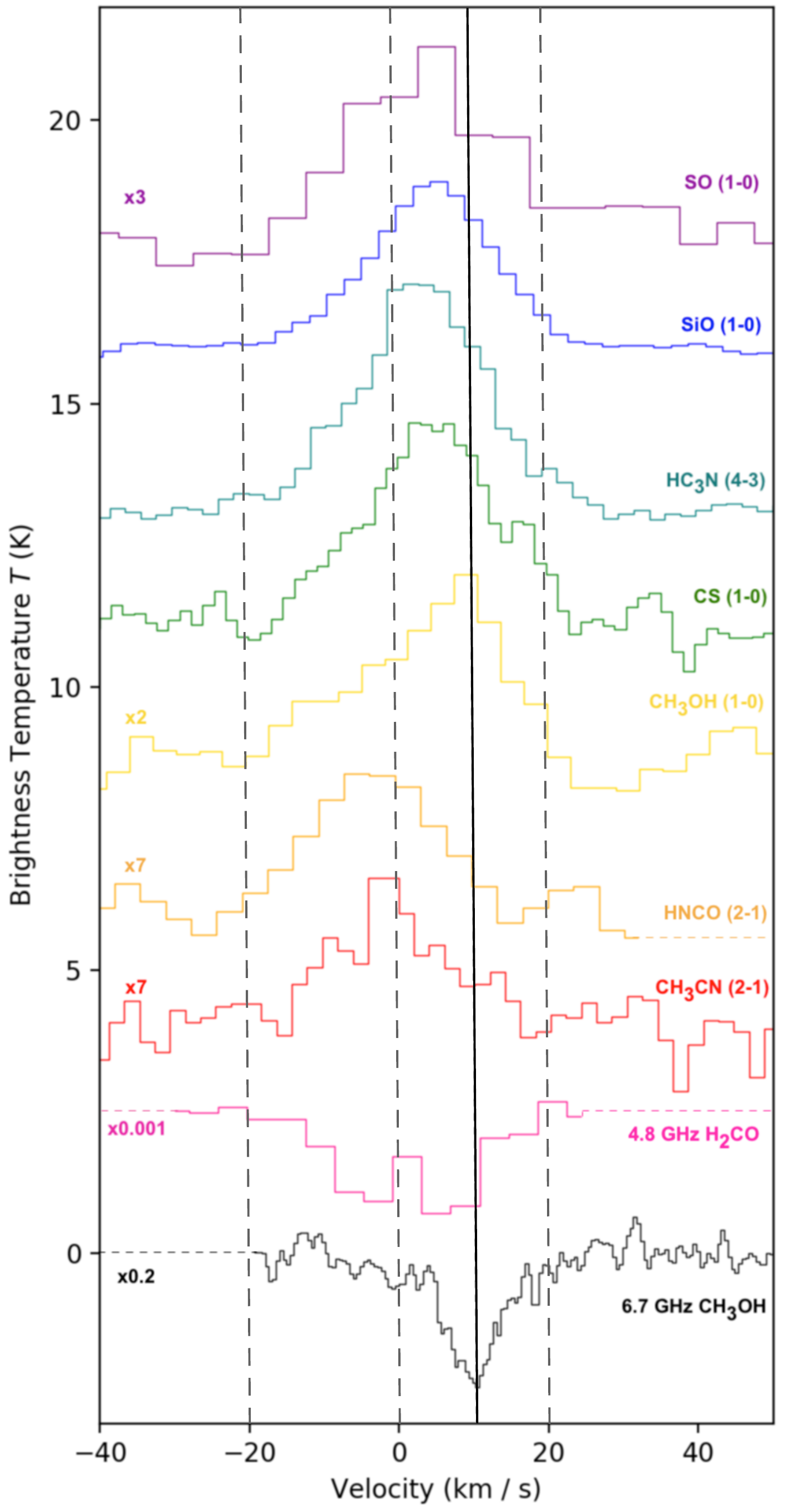} 
 \caption{Comparison of the 6.7 GHz \meth~absorption line (black spectrum) and the 4.8 GHz \form~absorption (pink spectrum) to seven molecular emission lines presented in \cite{dom16} (see their Table 2 and Figure 6). All spectra were integrated over the same region (3\arcsec)~aperture, centered on N3: $\alpha$(J2000)=17\h46\m21\fs0, ~$\delta$(J2000)= --28\degr50\arcmin03\farcs9. Each spectrum is labeled with its transition on the right side. The six spectral lines that have been vertically scaled, for comparison purposes, are indicated on the left side by the multiplied value. The vertical {solid black} line shows the central velocity of the \meth~absorption line (Table \ref{params}). {The three vertical grey dashed lines are reference lines at $-$20, 0, and +20 \kms.}  }
\label{comparison}
\end{figure}

\section{Results and Discussion}
\label{dis}

\subsection{6.7 GHz \meth~Absorption Feature}
\label{abs}

\begin{figure*}[tb!]
\centering
\includegraphics[scale=1.56]{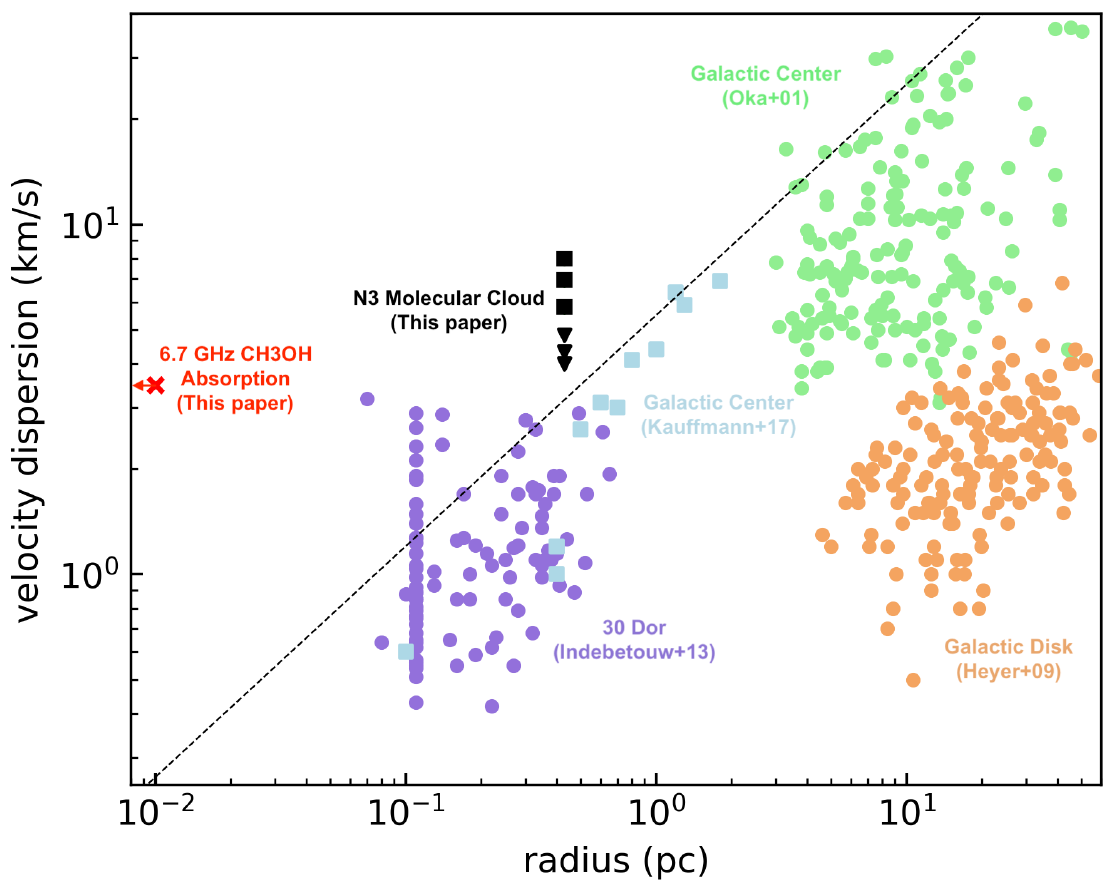} 
 \caption{The Larson linewidth-size ($\sigma$$-$$R$) relationship of molecular clouds. The red `x' shows the absorption feature presented in Section \ref{abs} and discussed in Section \ref{kin}. The black squares and triangles show the kinematics of the N3 molecular cloud for Components 1 and 2, respectively (see Table \ref{whole}). The dashed line shows the expected linewidth-size relationship for Galactic Center clouds from \cite{kauffmann17a} (see Equation \ref{e1}). The light blue squares are from \cite{kauffmann17a}. The light green data points are from \cite{oka01a} for gas clouds in the Galactic Center. The orange data points are from \cite{heyer09} showing the Galactic Disk cloud kinematics. The purple data points are from \cite{remy13} showing measurements from 30 Doradus in the Large Magellanic Cloud.  }
\label{graph}
\end{figure*}

The inset in Figure \ref{finder} left shows the 6.7 GHz \meth~absorption in the 9.97 \kms~channel. The feature is shown to be spatially aligned with the point-source N3 (red contour in inset), suggesting the absorbed continuum emission is likely from N3. We extracted the \meth~absorption using a region size comparable to the point spread function of the observations {(11\arcsec~$\times$ 5\arcsec~aperture)}, with the resulting spectrum presented in Figure \ref{finder}, right. We fit the spectrum in Figure \ref{finder}, right, {with a single Gaussian component}, using the python program \textit{pyspeckit} \citep{2011ascl.soft09001G}, and report the central velocity, velocity dispersion, and depth of the fit profile in Table \ref{params}. 

The measured central velocity of the 6.7 GHz \meth~absorption (9.94 \kms; vertical dashed line in Figure \ref{comparison}) is comparable to molecular emission velocities detected in the N3 Molecular Cloud \citep[$-$13 to + 25 \kms;][]{dom16}. Figure \ref{comparison} shows a comparison of the \meth~absorption profile (from Figure \ref{finder}, right) to molecular gas emission from seven transitions in \cite{dom16} and the \form~absorption line presented in Section \ref{obs}. The velocity range of the 6.7 GHz \meth~absorption feature, $+5$ to $+15$ \kms, overlaps with emission from all seven transitions (see Figure  \ref{comparison}). However, only the \meth~(1$_0$$-$0$_0$) line (yellow spectrum) peaks at the same velocity (\til9 \kms). All other emission lines have peak emission velocities lower than the 6.7 GHz \meth~absorption line, with central velocities ranging from $-5$ to $+5$ \kms. 

Similar to the \meth~absorption feature, the 4.8 GHz \form~absorption is also point-like and centered on N3. However, unlike the \meth~absorption feature, the \form~absorption contains two velocity components centered roughly at $-$3 and +8 \kms~(Figure \ref{comparison}). Further, the large brightness temperatures in both absorption components (\til$-$1500 K; the brightness temperature of the \form~spectrum in Figure \ref{comparison} was divided by 10$^3$ for comparison purposes), and spatial overlap with N3, suggests the absorbed continuum is likely from N3 and not the Cosmic Microwave Background.

\subsection{Line-of-sight Arrangement}
\label{arrange}

The detection of both \meth~and \form~absorption features indicates some of the molecular gas is located on the near-side of the continuum-emitting source N3. 
{This absorption detection opens back up a possibility ruled out in \cite{dom16}. \cite{dom16} did not detect any absorption features in their spectral line study and therefore concluded the cloud is located behind N3. Further, their conclusion places N3 in the GC and therefore they ruled out the possibility that N3 is a background AGN. However, the detection of absorption features in our data indicates some of the gas is located on the near-side of N3 and therefore there are no physical arguments that exclude an AGN. }
Further, the broken power-law spectrum of N3 is also consistent with SEDs of other AGN \citep{dom16}. Therefore, the possibility that N3 is a background AGN should remain viable. 

\subsection{CMZ Kinematics on Small Size Scales}
\label{kin}

\begin{figure}[tb!]
\includegraphics[scale=0.6]{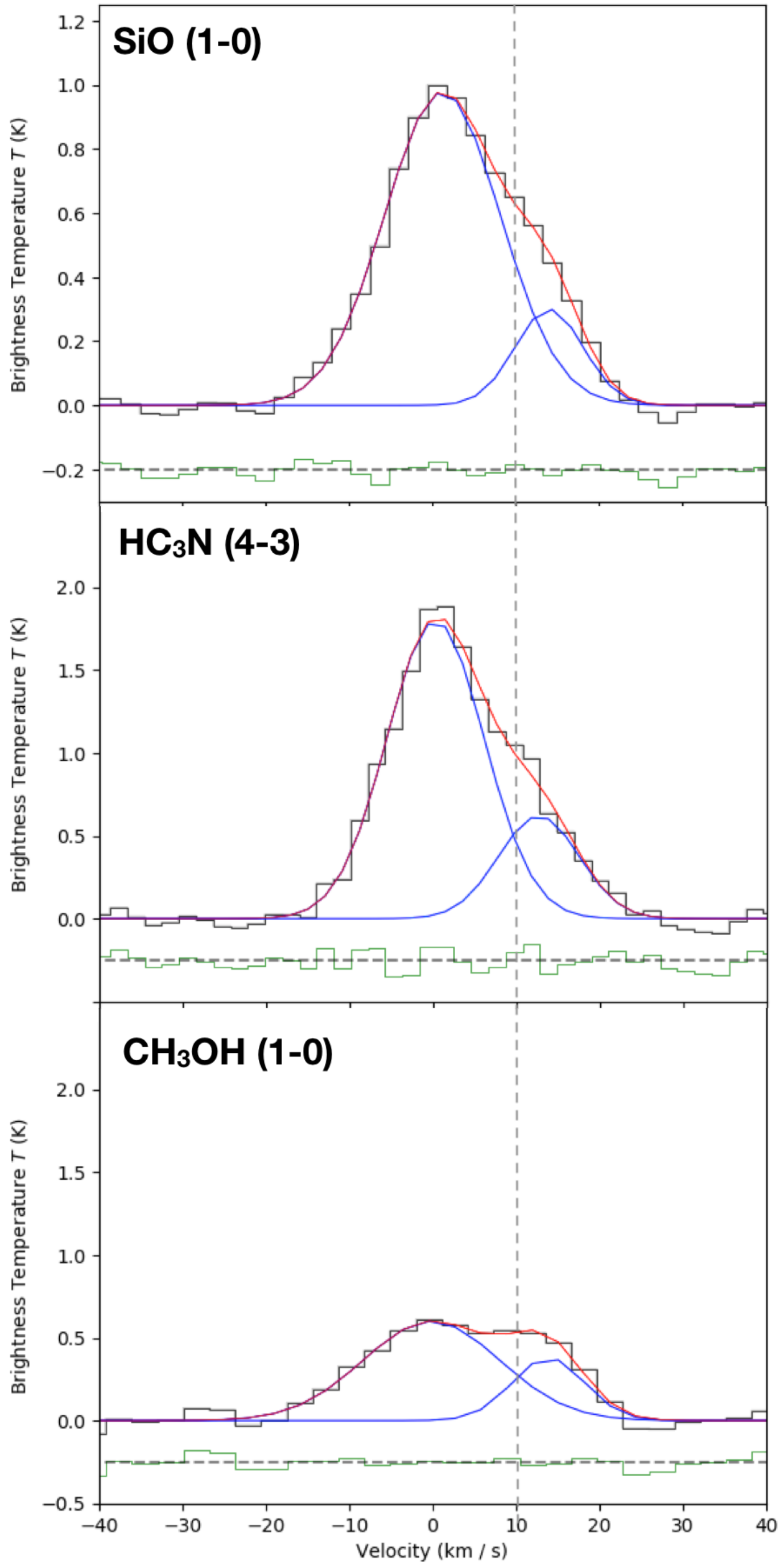}
 \caption{Black histogram shows the spectral profiles of SiO (1$-$0), \cyano (4$-$3), and \meth~(1$-$0), 
 from \cite{dom16},  integrated over the whole cloud ($R$=11\farcs2=0.43~pc; centered at \ra=17\h46\m20\fs8, \dec=-28\degree50\arcmin01\arcsec). In all three panels the spectra are best fit with two velocity components (blue Gaussians). The fitted parameters of each component for each transition are listed in Table \ref{whole}. The red line shows the combined fit of the two components, with the green histogram showing the residuals of the fit. The vertical grey dashed line shows the central velocity of the 6.7 GHz \meth~absorption feature.  }
\label{fig4}
\end{figure}

The 6.7 GHz \meth~absorption feature has an unusually large velocity dispersion, 3.8 \kms, for the size-scale of the continuum source N3. The Q-band observations in \cite{dom16}  suggest an upper limit to its angular size of 0\farcs26, implying a maximum size-scale of \til2000 AU, or $\leq$0.01 pc, at a GC distance of 8 kpc. Figure \ref{fig4}, shows the linewidth-size relationship of clouds from multiple studies compared to the measured linewidth-size of the \meth~absorption feature (red X).

We can compare our measured velocity dispersion to the predicted velocity dispersion, for a 0.01 pc cross-section size scale, using the following equation from \cite{kauffmann17a}: 
\begin{equation}
\sigma(v)=(5.5 \pm 1.0)~\textrm{km~s}^{-1}~(r_{e\!f\!f}/\textrm{pc})^{0.66 \pm 0.18}
\label{e1}
\end{equation}
where $r_{e\!f\!f}$ is the effective radius of the region and $\sigma(v)$ is the velocity dispersion. This relationship is shown as the dashed line in {Figure \ref{graph}}. For a 0.01 pc size scale the predicted velocity dispersion is 0.26$^{+0.17}_{-0.09}$ \kms, roughly an order of magnitude \textit{lower} than our measurements. Additionally, the measured velocity dispersion is also larger then the estimated sound speed of the gas, 0.6 \kms, using a gas temperature of \til100 K \citep{dom16} and equation 1 from \cite{kauffmann17a}. This estimated sound speed value suggests the N3 molecular cloud has a Mach number of 6, indicating the large linewidth is $not$ produced by thermal broadening. 
{This Mach number for the N3 cloud is comparable to other clouds in the GC which have a wide range of Mach numbers from 1.5 to 50 \citep[\eg,][]{mills17a, kauffmann17a}.}
The detection of a broad spectral line across such a small size scale could indicate that: 1) we are measuring turbulence along the depth of the cloud, 2) we are observing a coherently rotating cloud off-axis, or 3) N3 is a localized source that is producing the broader linewidth. We examine all three scenarios in the following sections.

\subsubsection{Scenario 1: Turbulence along the depth of the N3 cloud}
\label{s1}

\begin{table}[bt!]
\caption{\textbf{Kinematics of the N3 molecular cloud}}
\centering
\begin{tabular}{ccc}
\hline\hline
\textbf{Parameter}\footnote{Where $v_c$ is the central velocity of the component, $\sigma$ is the velocity dispersion and $T_{B}$ is the peak brightness temperature.} & \textbf{Component 1~~~~}  & \textbf{Component 2} \\
\hline
\multicolumn{3}{l}{\textbf{SiO (1$-$0) }} \\ [0.05cm]
\hline
$v_{c}$  	&  1.3 $\pm$ 0.7 \kms 	& 14.1 $\pm$ 1.2 \kms \\
$\sigma$ 	& 7.0 $\pm$ 0.6 \kms	&  4.0 $\pm$ 1.0 \kms   \\
$T_{B}$   	&  0.98 $\pm$ 0.03 K	& 0.30 $\pm$ 0.08 K  \\
\hline
\multicolumn{3}{l}{\textbf{\cyano~(4$-$3) }} \\ [0.05cm]
\hline
$v_{c}$ 	& 0.3 $\pm$ 0.5 \kms	& 12.7 $\pm$ 1.2 \kms   \\
$\sigma$ 	&  5.8 $\pm$ 0.4 \kms 	& 4.8 $\pm$ 0.8 \kms  \\
$T_{B}$  		& 1.80 $\pm$ 0.06 K 	& 0.62 $\pm$ 0.09 K  \\
\hline
\multicolumn{3}{l}{\textbf{\meth~(1$-$0) }} \\ [0.05cm]
\hline
$v_{c}$ 	& -0.1 $\pm$ 1.2 \kms 	& 13.9 $\pm$ 0.8 \kms  \\
$\sigma$ 	&  8.0 $\pm$ 1.0 \kms 	& 4.3 $\pm$ 0.8 \kms  \\
$T_{B}$  		& 0.60 $\pm$ 0.03 K	& 0.38 $\pm$ 0.09 K   \\ 
\hline\hline
%
\end{tabular}
\label{whole}
\end{table}




One possible scenario for the difference between the measured and predicted velocity dispersion values of the \meth~absorption feature is that we are measuring the velocity dispersion across the depth of the cloud, where the thickness of the cloud is larger than the observed cross-section of the point source. If the velocity dispersion is caused by turbulence across the thickness of the cloud, then the velocity dispersion of the pencil beam through the cloud is a 1-D profile. Therefore, the 3-D velocity dispersion would be $\sqrt{3}$ times larger than the measured 1-D velocity profile, assuming the turbulence is isotropic. In this scenario, the 3-D velocity dispersion, for a 3.84 \kms~1-D velocity profile, is 6.58 \kms. Using Equation \ref{e1}, this velocity dispersion corresponds to a size scale of 1.3$^{+1.0}_{-0.3}$ pc. The observed diameter of the N3 molecular cloud is \til10$-$25\arcsec~{\citep[0.4$-$1.0 pc;][]{dom16}}, indicating the detected velocity dispersion is roughly comparable for the size-scale of the whole N3 molecular cloud. To test this idea we investigate the kinematics of the entire N3 molecular cloud. 

Figure \ref{fig4} shows the spatially averaged SiO (1$-$0), \cyano~(4$-$3), and \meth~(1$-$0) \citep[data from][see their Figure 6]{dom16} emission spectrum integrated over the whole cloud ($r_{eff}$ = 11\farcs2 = 0.43 pc). These spectra were also fit using the python program $pyspeckit$. The properties of the best-fit Gaussian components are listed in Table \ref{whole}. 
{All three spectra} show two velocity components in the N3 cloud; a low velocity component around 0 \kms~(Component 1) and a high velocity component around 13 \kms~(Component 2). The central velocity of Component 2 is comparable to that of the \meth~absorption feature, indicating the absorption could be associated with this gas component. The velocity dispersion of both components, for all 3 transitions, are plotted on Figure \ref{graph}. In general, the N3 cloud appears to have a higher velocity dispersion then other GC clouds on similar 0.5-1.0 pc size scales. Our measured velocity dispersion of the \meth~absorption feature is comparable to the velocity dispersion across the entire cloud. Therefore, the broad linewidth could be produced by turbulence along the depth of the cloud, where the N3 cloud has a thickness of roughly 1 pc at this location.

\begin{figure}[tb!]
\centering
\includegraphics[scale=0.3]{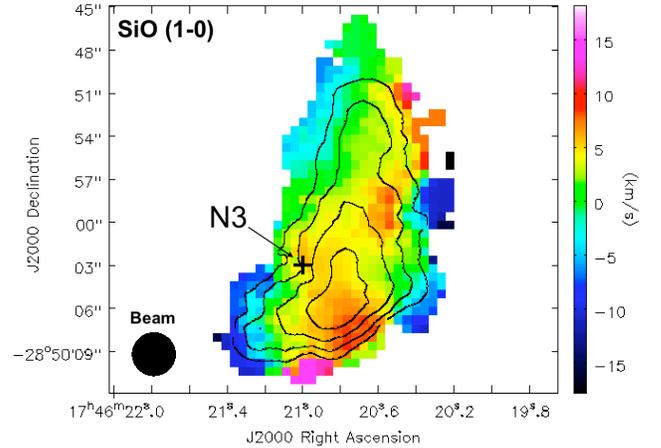} 
 \caption{Intensity weighted velocity distribution (moment 1) of the N3 molecular cloud in SiO (1$-$0) emission from \cite{dom16}. Contours show the integrated intensity (moment 0) of the SiO (1$-$0) at 20, 40, 60 and 80$\%$ of the peak intensity (1.184 \jybe~\kms). } 
\label{mom1}
\end{figure}

\subsubsection{Scenario 2: Large-scale cloud rotation}
\label{s3}

Another scenario that could cause the large velocity dispersion detected along a pencil beam through the cloud is a large scale cloud rotation. Observing a coherently rotating cloud off-axis could produce broad lines by blending multiple velocity components through the rotating cloud. To investigate this idea we consider the intensity-weighted velocity distribution (moment 1) across the cloud. If the cloud were rotating we would expect to see a velocity gradient across the cloud as a signature of rotation. 

Figure \ref{mom1} shows the intensity-weighted velocity distribution (moment 1) across the cloud in the SiO (1$-$0) transition from \cite{dom16}. Most of the lower velocity emission (below 0 \kms) is located towards the edge of the cloud, with the higher velocity emission (above 0 \kms) located towards the center of the cloud. There does not appear to be a large-scale velocity gradient across the entire cloud but rather a gradient from the `outside' inwards. The absence of a large scale velocity gradient across the cloud suggests that the large velocity dispersion is \textit{not} due to cloud rotation.

\subsubsection{Scenario 3: N3 embedded in the molecular cloud}
\label{s2}

Lastly, a broad absorption feature could be produced if N3 is $embedded$ in the molecular cloud, assuming N3 is located in the GC. In this case the compact-source would be enveloped in molecular material and surrounded by the two velocity components, discussed in Section \ref{s1}, which could represent the near and far sides of the cloud. If N3 was a massive compact object the broad linewidth on the 0.01 pc size scale could be caused by the gravitational interaction between the gas and N3. Additionally, the morphology of the cloud, which appears to wrap around N3 (\eg, see Figure \ref{mom1}), may support this claim.  

However, this described scenario disagrees with our \form~observations presented in Section \ref{obs}. 
The two \form~absorption components are comparable to the two velocity components averaged over the entire cloud (see discussion in Section \ref{s1}, Figure \ref{fig4}, and Table \ref{whole}). 
The detection of absorption in both velocity components is consistent with the cloud being located in front of the N3 point source. The lack of a {distinct} secondary component in the \meth~transition could be due to differences in the gas properties between the two velocity components {(\eg, temperature)}. 
{The residuals of the \meth~fitted spectrum are consistently negative between $-$10 and 0 \kms, hinting there could be a secondary component at this velocity range. However, the depth of this feature is at the 1$\sigma$ level (\til2 K). Higher sensitivity 6.7 GHz \meth~observations are needed to determine if this feature is a secondary absorption component. }
{Additional} follow up observations at high spectral resolution in multiple transitions are {also} necessary to investigate possible causes for the variation in the gas properties in the two components of the cloud. 


\section{Summary}

We present the detection of 6.7 GHz \meth~and 4.8 GHz \form~absorption towards the `N3' point-source (see Section \ref{abs}). The N3 point-source is located towards the Galactic Center Radio Arc and is unresolved at an angular  size of 0\farcs26 \citep[$<$ 0.01 pc for a source in the Galactic Center,][]{dom16}. The detection of the absorption features indicates that some of the gas in the N3 molecular cloud is located \textit{in front} of the point source N3 (Section \ref{arrange}). We also compare the kinematics of this absorption feature to the Larson size-linewidth relationship (Section \ref{kin}). The Larson size-linewidth relationship suggests a velocity dispersion of 0.26 \kms~for a size scale of 0.01 pc. However, fitting the kinematics of the \meth~absorption feature indicates a velocity dispersion of 3.8 \kms, an order of magnitude higher then predicted. We argue this relatively broad velocity dispersion is caused by turbulence along the depth of the cloud, where the thickness of the cloud is larger then the cross-section of the absorption feature. {Analysis presented in Section \ref{s1} suggests the N3 cloud has a thickness of roughly 1 pc.} Further, the detection of 4.8 GHz \form~absorption towards both velocity components of the N3 cloud is consistent with N3 being located behind the entire cloud. Therefore, N3 could be a background AGN.


\section{Acknowledgements}

ATB would like to acknowledge the funding provided from the European Union's Horizon 2020 research and innovation programme (grant agreement No 726384). DL would like to thank Katana Colledge for her assistance in setting up the VLA observations used to make the \form~data cube.
We would also like to thank Dr.  Alexei Kritsuk and Dr. Juergen Ott for their helpful insight on these results. 
NB would also like to thank Dr. Drew Medlin at NRAO for assisting with pipeline calibration. 
{We would also like to thank the anonymous reviewer for their helpful comments on this work.}

\software{CASA \citep{2011ascl.soft07013I}; $pyspeckit$ \citep{2011ascl.soft09001G}}


\bibliographystyle{aasjournal}
\bibliography{N3.bib}

\end{document}